\documentclass[a4paper,11pt]{article}

\usepackage{amsmath,amssymb,amsfonts} % For math symbols and fonts
\usepackage{graphicx} % For including figures
\usepackage{hyperref} % For hyperlinks in the document
\usepackage{authblk} % For author affiliations
\usepackage[numbers,sort&compress]{natbib} % For citations
\usepackage{geometry} % For adjusting margins
\usepackage{subfig}

\title{Influence of Yttrium(Y) on properties of Lanthanum Cobalt Oxides($LaCoO_3$)}

\author[1]{Mohammad Abu Thaher Chowdhury}
\author[2]{Shumsun Naher Begum}
\affil[1,2]{\small  Condensed Matter Lab, Department of Physics, Shahjalal University of Science and Technology, Sylhet, 3114, Bangladesh}
\affil[1]{BCSIR, Dhaka, 1205,Bangladesh}

\date{August, 2019}  % Remove date

\begin{document}
\maketitle

\begin{abstract}
Many materials exhibit various types of phase transitions at different temperatures, with many also demonstrating polymorphism. Doping materials can significantly alter their conductivity. In light of this, we have investigated the electrical conductivity of $LaCoO_3$, specifically its temperature dependence when doped with Yttrium (Y). The crystal structure of Lanthanum Yttrium Cobalt oxide $(La_{1-x}Y_x Co O_3)$ adopts a perovskite form, characterized by the general stoichiometry $ABX_3$, where A and B are cations, and X is an anion. This material undergoes a magnetic phase transition between $50-100$ K, a structural phase transition between $100-300$ K, and an insulator-to-metal transition at $500$ K. At room temperature, $LaCoO_3$ exhibits polaron-type hopping conduction. Our aim was to understand the electrical conductivity at $300$ K and how it varies with temperature when $La^{3+}$ is replaced by $Y^{3+}$. The electrical properties of the perovskite crystal are consistent with small polaron hopping conduction, which theoretically follows Mott's variable range hopping model, where conductivity obeys an exponential law, and resistivity follows an inverse exponential pattern. In this work, we compare the experimental resistivity graph with the theoretical inverse of the conductivity graph, showing that our experimental results align with the polaron hopping conduction model within a certain range. Additionally, the experiment confirms polymorphism in various cases. We observed that increasing the concentration of $Y^{3+}$ enhances the metallic properties of $La_{1-x} Y_x Co O_3$, and we found a significant correlation between conductivity and symmetry. Furthermore, the study highlights the material's phase transitions and polymorphic behavior. %\lipsum[1]
\end{abstract}


%-------------------------------------------
% Paper Body
%-------------------------------------------

%--- Section ---%
\section{Introduction}

Perovskites~\citep{Wenk_Bulakh_2016}, was first discovered by Gustav Rose and named after Russian minerelogist L.A. Perovski, form a large and diverse family of isotropic crystalline ceramics, commonly the perovskite structure represented by the chemical formula $ABX_3$~\citep{Goldschmidt1926DieGD}, where $A$ and $B$ are cations of considerably different sizes, and $X$ is an anion. This highly adaptable $ABX_3$ perovskite crystal structure is distinguished by a three-dimensional network of corner-sharing $BX_6$ octahedra, with $A$-site cations filling the $12$-coordinate cavities created by the $BX_6$ network~\citep{Orlovskaya2004MixedIE}. The ability to substitute different cations at the $A$ and $B$ positions allows for the formation of complex perovskites, which may exhibit both ordered and disordered structural variants.

In our study, we focus on doping Yttrium ions $(Y^{3+})$ into the Lanthanum ion $(La^{3+})$ sites within $LaCoO_3$, a compound that has garnered considerable attention due to its unique magnetic properties. These properties are primarily attributed to the spin state of $Co^{3+}$ ions~\citep{Knížek2005}. At low temperatures, $LaCoO_3$ remains nonmagnetic because the $Co^{3+}$ ions are in a low spin (LS) state, characterized by six electrons in the $t_{2g}$ orbitals and vacant $e_g$ orbitals (LS, $t^6_{2g} e^0_g, S = 0$). However, between $50$ and $100 K$, $LaCoO_3$ undergoes a significant magnetic phase transition, which is driven by the thermal excitation of $Co^{3+}$ ions into either an intermediate spin (IS, $t^5_{2g}e^1_g, S = 1$) or high spin (HS, $t^4_{2g} e^2_g, S = 2$) state.

Keisuke Sato and colleagues~\citep{Sato_2009} have observed that as the temperature increases, the magnetization of $LaCoO_3$ also increases, peaking around $100 K$, after which it gradually decreases. A second notable transition, occurring between $500$ K and $550$ K, is identified as an insulator-to-metal (I-M) transition, which is accompanied by additional changes in the material’s paramagnetic properties. According to research by K. Knizek et al.~\citep{Knížek2005}, these magnetic and electronic transitions are also associated with subtle but significant changes in the crystal structure, which typically assumes a rhombohedral perovskite-type $R3c$ form.

Two primary effects have been documented in the literature: firstly, the ionic radius of $Co^{3+}$ increases from $r_{LS} = 0.545 \AA$ to $r_{IS} = 0.56 \AA$ or $r_{HS} = 0.6 \AA$; secondly, the $Co^{3+}$ ion in the IS or HS state becomes Jahn-Teller active~\citep{Jahn_1937}, which may induce distortions in the $CoO_6$ octahedra. Recent work by G. Maris and his team has revealed a monoclinic distortion of the $LaCoO_3$ structure (space group $I2/a$) along with a Jahn-Teller distortion of $CoO_6$ octahedra, as observed through single-crystal X-ray diffraction studies conducted between $100 $ K and $300$ K~\citep{Maris_2003}.

Interestingly, the analogous compound $YCoO_3$ also experiences a diamagnetic-paramagnetic transition, albeit in a more diffusive manner and at higher temperatures, ranging from $450$ K to $800$ K~\citep{Knizek_2006}. In $YCoO_3$, the I-M transition occurs around $750$ K~\citep{Knizek_2006}. Given these insights, our research is centered on investigating the effects of substituting $Y^{3+}$ for $La^{3+}$ in $LaCoO_3$, with a particular focus on understanding how this doping alters the electronic properties of the material.

%--- Section ---%
\section{Theoretical background}\label{sec2}
The electrical conductivity of perovskites is intricately linked to the structure of their composites, particularly in systems where small polarons are the primary charge carriers. In the context of our study, the perovskite sample exhibits conductivity behavior consistent with small polaron hopping, a mechanism that is well-described by the Mott's variable range hopping (VRH) model~\citep{Mott_1969}. This model is crucial for understanding temperature-dependent conduction in strongly disordered systems with localized charge carrier states.

Small polarons are quasiparticles that arise due to strong interactions between electrons and the lattice, leading to localized states~\citep{Devreese_2000,DEVREESE2005}. These polarons follow the hopping conduction mechanism, where charge carriers (electrons or holes) move between localized sites via thermally activated jumps. The hopping conduction model, particularly Mott’s VRH, explains that in such disordered systems, conduction occurs not through a band-like mechanism but through jumps from one localized state to another. This process is highly dependent on temperature, as it dictates the likelihood of electrons overcoming potential barriers between localized states.

In low-energy states, the wavefunctions of these electrons have large amplitudes, which means that electrons are more likely to be found in these low-energy regions due to the minimal overlap integrals between states~\citep{Pavarini_2017}. This creates a situation where electrons prefer to remain localized unless they acquire sufficient thermal energy to overcome the potential fluctuations that confine them. At finite temperatures, electrons can gain enough energy to hop between localized states, leading to the observable hopping conduction.

According to the hopping conduction model~\citep{Mott_1969,Yu_2004}, the resistivity $(\rho)$ of perovskites like $LaCoO_3$ and $La_{1-x}Y_x Co O_3$ should follow an exponential law, which is mathematically expressed as:
\begin{align}
    \rho &= \rho_0 e^{\frac{W}{k_B T}}
\end{align}

where $\rho$ represents the temperature-dependent resistivity, $\rho_0$ is the initial resistivity at room temperature, $W$ denotes the activation energy, $k_B$ is the Boltzmann constant, and $T$ is the temperature .

In addition to this exponential behavior, the resistivity is also expected to follow Mott’s variable range hopping law, characterized by a $\displaystyle{T^{-\frac{1}{4}}}$ temperature dependence~\citep{Mott_1969}. Moreover, in the regime described by Efros and Shklovskii (ES), variable range hopping in the presence of a Coulomb gap, the resistivity follows a $\displaystyle{T^{-\frac{1}{2}}}$ law~\citep{Efros_1975}. These relationships provide a robust framework for understanding the temperature-dependent electrical properties of perovskites and further illustrate the role of structural and compositional factors in determining their conductivity.

%--- Section ---%
\section{Experimental method}\label{sec3}
The synthesis of $LaCoO_3$ and $La_{1-x}Y_xCoO_3$ perovskites was carried out using the solid-state reaction method~\citep{Lide_2017}, a well-established technique for preparing ceramic materials. Initially, $5$ grams of Lanthanum Oxide $(La_2O_3)$ were thoroughly ground with $5$ grams of Cobalt Oxide $(Co_2O_3)$ to form a homogenous mixture. Yttrium Oxide $(Y_2O_3)$ was then added to the mixture in precise proportions corresponding to $x = 0.2, 0.4, 0.6, and 0.8$ for the yttrium $(Y)$ content, ensuring accurate doping levels. This mixture was further agitated in a mortar under acetone for $7$ hours to enhance the homogeneity of the powders, as acetone assists in reducing particle size and eliminating impurities.

Following the preparation of the mixtures, they were deposited into a crucible and subjected to pre-sintering at $800^{\circ} C$ for $5$ hours in a furnace under an inert atmosphere, which is essential to prevent oxidation and ensure the stability of the compounds. This step facilitates the initial formation of the perovskite phase. After pre-sintering, the resulting materials were reground and hand-milled for an additional $3$ hours under acetone to further refine the particle size and enhance the uniformity of the mixture.

The powders were then pressed into pellets using a hydraulic press, applying a constant pressure of $20 Pa$. This uniform pressure ensures consistent thickness across all samples, which is critical for reliable comparisons of their physical properties. The pellets were subsequently annealed at $900^{\circ} C$ for $5$ hours, with the temperature being ramped up at a rate of $50^{\circ} C$ per minute. The annealing process was followed by dehydration at $100^{\circ} C$ for $7$ hours to remove any residual moisture, which could otherwise interfere with the sintering process.

The final sintering was performed at $1200^{\circ} C$ for $5$ hours, a critical step that consolidates the material, improving its density and ensuring the formation of the desired perovskite phase. The chemical reactions that likely occur during pre-sintering and sintering are outlined below:
\begin{align*}
    3La_2O_3 + 2Co_3O_4 &\rightarrow 6 LaCoO_3\\
    3La_2O_3 + 3Y_2O_3 + 2Co_3O_4 &\rightarrow 6 LaYCoO_3 + 4O_2\uparrow
\end{align*}

These reactions result in the formation of pure $LaCoO_3$ and $La_{1-x}Y_xCoO_3$ perovskites, essential for their subsequent structural and electrical characterization.

To investigate the crystallographic structure of the synthesized perovskites, X-ray diffraction (XRD) analysis was conducted using a diffractometer equipped with CuK$\alpha$ monochromatic radiation (wavelength $\lambda = 0.154056 nm$), operating at a tube current of $10$ mA and a voltage of $30$ kV. Diffraction profiles were recorded over a Bragg angle range of $15^{\circ}$ to $85^{\circ}$, providing detailed insights into the crystal structure and phase purification of the perovskites.

Surface morphology analysis was performed using a scanning electron microscope (SEM) equipped with Ultra Dry Energy Dispersive X-ray (EDX) analysis, enabling the examination of microstructural features and elemental composition. For electrical property measurements, the samples were provided with electrical contacts, and the conductivity was derived from resistivity, current, and voltage measurements. These measurements were carried out using a Keithley $6517B$ Electrometer, which allowed precise determination of resistivity ($\rho$), current ($I$), and voltage ($V$) across different temperatures, providing critical data on the temperature-dependent electrical behavior of the perovskite films.

%--- Section ---%
\section{Result and Discussion}\label{sec4}

%--- Subsection ---%
\subsection{Electric Property}
This observation that the current magnitude increases with rising $Y^{3+}$ concentration aligns well with the theory of polarization quanta associated with small polaron hopping. The phenomenon can be further elucidated by examining the structural and electronic changes that occur within the perovskite lattice as a result of doping.

At room temperature, the $Co-O-Co$ bond angle in $YCoO_3$ is relatively small, measuring around $148^{\circ}$, and this angle remains stable up to $600$ K. Beyond this temperature, however, a spin transition is initiated, causing a decrease in the bond angle and leading to a slightly greater expansion of the $CoO_6$ octahedra compared to the overall lattice expansion. In contrast, the $Co-O-Co$ bond angle in $LaCoO_3$ at room temperature is $164^{\circ}$, and this angle increases with temperature, resulting in a smaller expansion of the $CoO_6$ octahedra relative to the lattice. The distinct structural behaviors between $YCoO_3$ and $LaCoO_3$ can be attributed to the higher symmetry present in the $LaCoO_3$ structure, which is more stable and less prone to releasing electrons than $YCoO_3$.

The difference in bond angles and expansion behaviors is significant because it highlights how the substitution of $La$ with $Y$ affects the perovskite structure and its electronic properties. $LaCoO_3$ exhibits a rhombohedral symmetry, which is more stable and less likely to distort under thermal stress. On the other hand, $YCoO_3$ has an orthorhombic structure, which is less symmetric and more susceptible to structural changes. The structural evolution from rhombohedral $LaCoO_3$ to orthorhombic $YCoO_3$ as the temperature increases can be traced back to the rotation of the cobalt-centered octahedra, a feature that drives the symmetry shift.

Neutron powder diffraction (NPD) studies of $YCoO_3$ at high temperatures (up to $1000$ K) confirm that the structure remains orthorhombic throughout this range. This persistent orthorhombic symmetry, driven by the rotation of the $CoO_6$ octahedra, contrasts with the more stable rhombohedral symmetry of $LaCoO_3$. The decrease in ionic radius as $Y$ is introduced into the lattice prompts a transition from higher symmetry structures (cubic or rhombohedral) to lower symmetry ones (orthorhombic). This structural distortion results in an increase in the number of free electrons, which, in turn, enhances the material’s electrical conductivity.

\begin{figure}[!t]
    \centering
    \includegraphics[scale = 0.85]{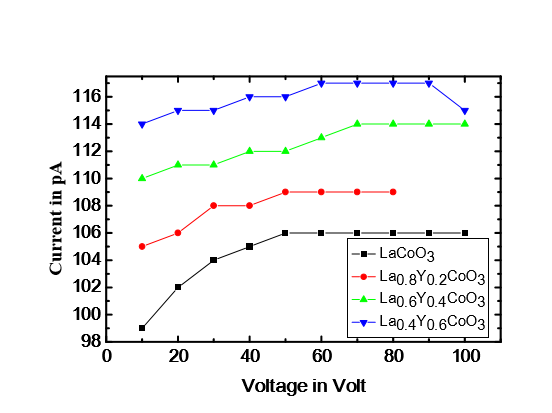}
    \caption{$I$ Vs $V$ curves for the samples $La_{(1-x)} Y_x CoO_3 ( x= 0, 0.2, 0.4, 0.6)$ at $300$ K}
    \label{fig:I-V}
\end{figure}

The relationship between symmetry and electronic configuration stability is clear: the higher symmetry of $LaCoO_3$ ensures greater stability for its electronic configuration, leading to less electronic activity and, therefore, lower conductivity compared to $YCoO_3$. As $Y$ concentration increases, the symmetry of the crystal decreases, and the structure becomes more conducive to electron movement, which explains the observed increase in current until a saturation point is reached at a particular voltage (see Fig.~\ref{fig:I-V}).

\begin{figure}[!b]
    \centering
    \includegraphics[scale = 0.5]{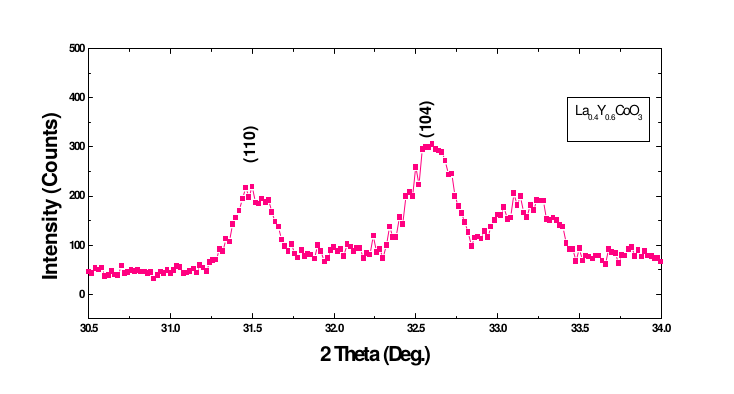}
    \caption{Profiles of ($I-2\theta$) scan of (110) and ($104$) Bragg diffraction of $La_{0.4} Y_{0.6} CoO_3$}
    \label{fig:structure}
\end{figure}

X-ray diffraction (XRD) patterns at room temperature for $x=0.6$ reveal that the peak, which is singular at $x=0$, splits into three distinct peaks in Fig.~\ref{fig:structure}, indicating a transition from rhombohedral to orthorhombic symmetry. However, this peak splitting was not observed in other concentrations, suggesting that the crystal symmetry decreases as the $Y^{3+}$ concentration increases. This reduction in symmetry correlates with the enhanced electrical conductivity, as the distorted structure allows for more effective small polaron hopping, thereby increasing the overall current.

%--- Subsection ---%
\subsection{Temperature dependent resistivity}\label{sec5}

\begin{figure}[!b]
    \centering
    \includegraphics[width = \textwidth]{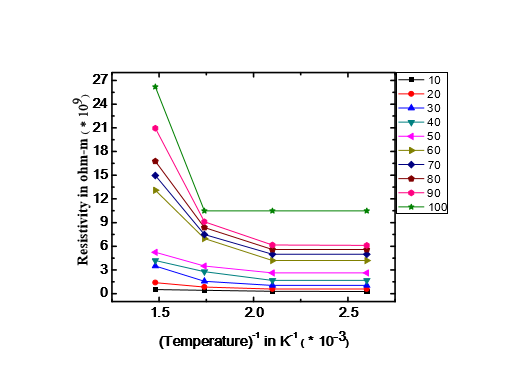}
    \caption{ Resistivity vs $(Temperature)^{-1}$ curve for sample $La_{0.8}Y_{0.2}CoO_3$ with voltage range $10-100$ V and with temperature ($T= 373$ K$,\ 473$ K$ ,\ 573$ K$ ,\ 673$ K)  }
    \label{fig:rho_T}
\end{figure}

\begin{figure}[t!]
        \subfloat[x = 0.0]{%
            \includegraphics[width=.48\linewidth]{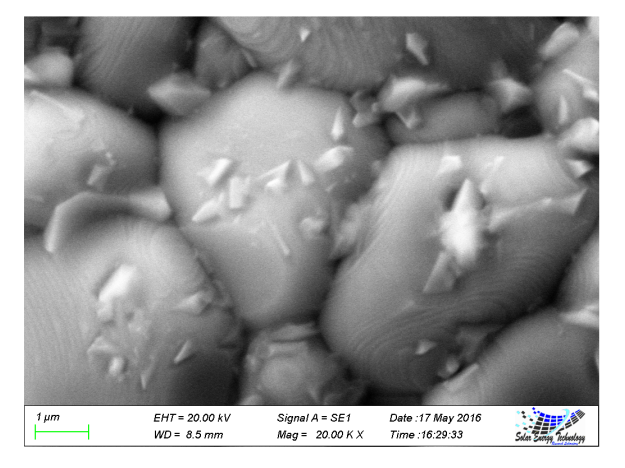}%
            \label{subfig:a}%
        }\hfill
        \subfloat[x = 0.2]{%
            \includegraphics[width=.48\linewidth]{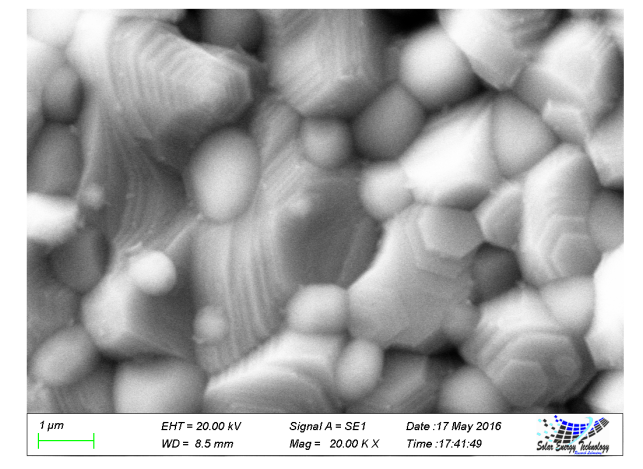}%
            \label{subfig:b}%
        }\\
        \subfloat[x = 0.4]{%
            \includegraphics[width=.48\linewidth]{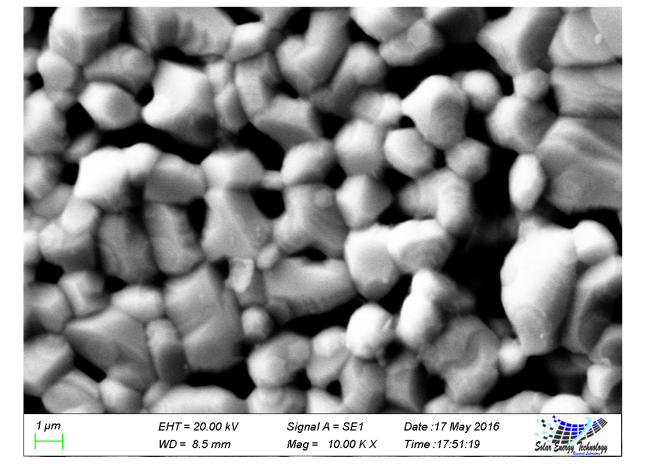}%
            \label{subfig:c}%
        }\hfill
        \subfloat[x = 0.6]{%
            \includegraphics[width=.48\linewidth]{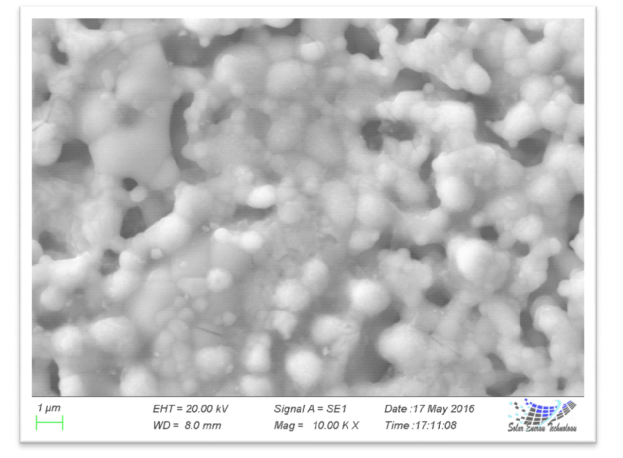}%
            \label{subfig:d}%
        }
        \caption{SEM pattern of $La_{1-x} Y_x CoO_3 ( x= 0, 0.2, 0.4, 0.6)$ was taken by high voltage $20$kV and magnified $10.00$ K times within the scale bar of $1$ $\mu$m.}
        \label{fig:sem}
    \end{figure}

Based on our experimental data, after careful analysis and comparison with theoretical predictions, we confirm that the system does follow an exponential law, although it deviates slightly from Mott's variable range hopping (VRH) model (which needs more experimental data to be sure). Specifically, after a certain temperature range, the resistivity behavior begins to plateau, indicating a voltage-dependent relationship rather than strictly adhering to Mott's VRH formula. This plateau suggests that while the VRH model offers a strong foundation for understanding the conduction mechanisms, the presence of external factors such as voltage may introduce deviations that are not fully captured by the model.

The gradual increase in the metallic properties of the compound with the incremental addition of $Y^{3+}$ aligns with previous findings, such as those reported by Ashwin V. et al., who observed that $LaCoO_3$ undergoes an insulator-to-metal (I-M) phase transition at temperatures between $500-550$ K~\citep{AswinV_2015}. Our experimental results further support this, demonstrating that as the temperature exceeds $573$ K, there is a noticeable increase in resistivity within the voltage range of $10-50$ V. However, a sharp rise in resistivity is observed within the voltage range of $60-70$ V, strongly indicating the occurrence of an I-M phase transition in Fig.~\ref{fig:rho_T}. This sharp increase confirms the onset of the I-M transition, as predicted by the theoretical framework.

The theoretical expression for resistivity in section \ref{sec2} can be expressed as $$\rho = \rho_0\frac{T}{T_0}\  exp\left[\frac{W}{K_B T}\right]$$ provides a valuable tool for understanding the temperature dependence of resistivity in these doped perovskites. Here, $T_0$ represents the initial temperature, and further data collection and analysis across various samples will allow us to refine this model and validate its applicability across a broader range of experimental conditions.

Moreover, observations from the scanning electron microscope (SEM) patterns [Figure \ref{fig:sem}] reveal that as the concentration of $Y^{3+}$ increases, there is a corresponding decrease in the grain size of the particles. This grain size reduction can be attributed to the smaller ionic radius of $Y^{3+}$ compared to $La^{3+}$, which affects the microstructural evolution of the perovskite material. The reduction in grain size, coupled with the increase in $Y^{3+}$ concentration, also leads to a rise in the number of grain boundaries within the material. Consequently, as free electrons migrate due to thermal excitation, they encounter more grain boundaries, leading to increased scattering events. This scattering contributes to the overall rise in resistivity as the temperature increases, consistent with the observed experimental results.

This intricate interplay between microstructural changes, doping concentration, and electrical properties underscores the complexity of conducting behavior in doped perovskites, and highlights the need for further detailed studies to fully unravel these phenomena.

%--- Section ---%
\section{Conclusion}\label{sec:conclusion}

Based on our comprehensive experimental and theoretical investigations, we have conclusively established that $La_{1-x}Y_x Co O_3$ exhibits notable polymorphism and undergoes an insulator-to-metal (I-M) phase transition within the temperature range of $500$ k$-600$ K. This polymorphic behavior is indicative of the compound's ability to adopt multiple structural forms depending on temperature and composition, a characteristic feature of perovskite materials subjected to various doping levels and thermal conditions.

Our findings reveal that the metallic properties of the perovskite increase with the rising concentration of yttrium. This enhancement in metallic character is directly related to the decrease in crystal symmetry as yttrium is substituted for lanthanum. As the yttrium content increases, the perovskite structure transitions from a higher symmetry to a lower symmetry phase, which facilitates greater electron mobility and contributes to the observed rise in metallic properties.

However, this increase in metallic behavior is accompanied by an increase in resistivity with temperature. This temperature-dependent resistivity rise indicates that, despite the enhanced metallic characteristics, the material's ability to conduct electricity diminishes at higher temperatures. This counterintuitive result can be attributed to the increasing scattering of charge carriers at elevated temperatures, which is consistent with the changes in crystal symmetry and structure induced by yttrium doping.

Additionally, while the system adheres to an exponential law for resistivity over certain temperature ranges, it deviates from Mott's variable range hopping (VRH) formula after a specific temperature threshold. This deviation suggests that, although the exponential relationship captures the general trend of resistivity, the detailed behavior of the conductivity in $La_{1-x}Y_x Co O_3$ does not fully conform to the VRH model. Instead, the observed resistivity behavior reflects a more complex interplay between structural changes, doping effects, and temperature variations.

In summary, the data support the notion that $La_{1-x}Y_xCoO_3$ undergoes significant structural and electronic transformations with increasing yttrium concentration. The material's polymorphic nature and the associated I-M phase transition underscore the intricate relationship between doping, crystal symmetry, and electronic properties, providing valuable insights into the behavior of doped perovskites in varying thermal and compositional conditions.

%-------------------------------------------
% Optional Contents
%-------------------------------------------

%--- Section ---%
\section*{Author Contributions}
\noindent\textbf{Mohammad Abu Thaher Chowdhury}: Conceptualization, Preparing sample, data collection using XRD and scanning electric microscope, Visualization and investigation using data, writing-original draft \\ \textbf{Shumsun Naher Begum}: Supervision.

%--- Section ---%
\section*{Funding}
This research was supported by Kanazawa University, Japan.

%--- Section ---%
\section*{Acknowledgments}
We are thankful to our colleagues Md Shafiul Alam Sajib and Assist. Professor Md. Omar Faruk who provided the expertise that greatly assisted the research, although they may not agree with all of the interpretations provided in this paper.

We have to express our appreciation to the Bangladesh Council of Scientific and Industrial Research for sharing their pearls of wisdom and instruments with us during the course of this research. We are also immensely grateful to Md. Gofur Ahmed for his comments on an earlier version of the manuscript, although any errors are our own and should not tarnish the reputations of these esteemed professionals.

\bibliographystyle{plainnat} % Choosing a bibliography style
\bibliography{references} % references.bib is the name of the .bib file

\begin{thebibliography}{16}
\providecommand{\natexlab}[1]{#1}
\providecommand{\url}[1]{\texttt{#1}}
\expandafter\ifx\csname urlstyle\endcsname\relax
  \providecommand{\doi}[1]{doi: #1}\else
  \providecommand{\doi}{doi: \begingroup \urlstyle{rm}\Url}\fi

\bibitem[Cho et~al.(2017)Cho, Uddin, and Alaboina]{Lide_2017}
S.-J. Cho, M.-J. Uddin, and P.~Alaboina.
\newblock Chapter three - review of nanotechnology for cathode materials in batteries.
\newblock In Lide~M. Rodriguez-Martinez and Noshin Omar, editors, \emph{Emerging Nanotechnologies in Rechargeable Energy Storage Systems}, Micro and Nano Technologies, pages 83--129. Elsevier, Boston, 2017.
\newblock ISBN 978-0-323-42977-1.
\newblock \doi{https://doi.org/10.1016/B978-0-323-42977-1.00003-0}.
\newblock URL \url{https://www.sciencedirect.com/science/article/pii/B9780323429771000030}.

\bibitem[{Devreese}(2000)]{Devreese_2000}
J.~T. {Devreese}.
\newblock {Polarons}.
\newblock \emph{arXiv e-prints}, art. cond-mat/0004497, April 2000.
\newblock \doi{10.48550/arXiv.cond-mat/0004497}.

\bibitem[Devreese(2005)]{DEVREESE2005}
J.T. Devreese.
\newblock Electron–phonon interactions and the response of polarons.
\newblock In Franco Bassani, Gerald~L. Liedl, and Peter Wyder, editors, \emph{Encyclopedia of Condensed Matter Physics}, pages 99--109. Elsevier, Oxford, 2005.
\newblock ISBN 978-0-12-369401-0.
\newblock \doi{https://doi.org/10.1016/B0-12-369401-9/00664-1}.
\newblock URL \url{https://www.sciencedirect.com/science/article/pii/B0123694019006641}.

\bibitem[Efros and Shklovskii(1975)]{Efros_1975}
A~L Efros and B~I Shklovskii.
\newblock Coulomb gap and low temperature conductivity of disordered systems.
\newblock \emph{Journal of Physics C: Solid State Physics}, 8\penalty0 (4):\penalty0 L49, feb 1975.
\newblock \doi{10.1088/0022-3719/8/4/003}.
\newblock URL \url{https://dx.doi.org/10.1088/0022-3719/8/4/003}.

\bibitem[Goldschmidt(1926)]{Goldschmidt1926DieGD}
Victor Goldschmidt.
\newblock Die gesetze der krystallochemie.
\newblock \emph{Naturwissenschaften}, 14:\penalty0 477--485, 1926.
\newblock URL \url{https://api.semanticscholar.org/CorpusID:33792511}.

\bibitem[{Jahn} and {Teller}(1937)]{Jahn_1937}
H.~A. {Jahn} and E.~{Teller}.
\newblock {Stability of Polyatomic Molecules in Degenerate Electronic States. I. Orbital Degeneracy}.
\newblock \emph{Proceedings of the Royal Society of London Series A}, 161\penalty0 (905):\penalty0 220--235, July 1937.
\newblock \doi{10.1098/rspa.1937.0142}.

\bibitem[Kn\'{\i}\ifmmode~\check{z}\else \v{z}\fi{}ek et~al.(2006)Kn\'{\i}\ifmmode~\check{z}\else \v{z}\fi{}ek, Jir\'ak, Hejtm\'anek, Veverka, Mary\ifmmode~\check{s}\else \v{s}\fi{}ko, Hauback, and Fjellv\aa{}g]{Knizek_2006}
K.~Kn\'{\i}\ifmmode~\check{z}\else \v{z}\fi{}ek, Z.~Jir\'ak, J.~Hejtm\'anek, M.~Veverka, M.~Mary\ifmmode~\check{s}\else \v{s}\fi{}ko, B.~C. Hauback, and H.~Fjellv\aa{}g.
\newblock Structure and physical properties of $\mathrm{Y}\mathrm{Co}{\mathrm{o}}_{3}$ at temperatures up to $1000\phantom{\rule{0.3em}{0ex}}\mathrm{K}$.
\newblock \emph{Phys. Rev. B}, 73:\penalty0 214443, Jun 2006.
\newblock \doi{10.1103/PhysRevB.73.214443}.
\newblock URL \url{https://link.aps.org/doi/10.1103/PhysRevB.73.214443}.

\bibitem[Kn{\'i}{\v{z}}ek et~al.(2005)Kn{\'i}{\v{z}}ek, Jir{\'a}k, Hejtm{\'a}nek, Veverka, Mary{\v{s}}ko, Maris, and Palstra]{Knížek2005}
K.~Kn{\'i}{\v{z}}ek, Z.~Jir{\'a}k, J.~Hejtm{\'a}nek, M.~Veverka, M.~Mary{\v{s}}ko, G.~Maris, and T.~T.~M. Palstra.
\newblock Structural anomalies associated with the electronic and spin transitions in lncoo3.
\newblock \emph{The European Physical Journal B - Condensed Matter and Complex Systems}, 47\penalty0 (2):\penalty0 213--220, Sep 2005.
\newblock ISSN 1434-6036.
\newblock \doi{10.1140/epjb/e2005-00320-3}.
\newblock URL \url{https://doi.org/10.1140/epjb/e2005-00320-3}.

\bibitem[Maris et~al.(2003)Maris, Ren, Volotchaev, Zobel, Lorenz, and Palstra]{Maris_2003}
G.~Maris, Y.~Ren, V.~Volotchaev, C.~Zobel, T.~Lorenz, and T.~T.~M. Palstra.
\newblock Evidence for orbital ordering in ${\mathrm{lacoo}}_{3}$.
\newblock \emph{Phys. Rev. B}, 67:\penalty0 224423, Jun 2003.
\newblock \doi{10.1103/PhysRevB.67.224423}.
\newblock URL \url{https://link.aps.org/doi/10.1103/PhysRevB.67.224423}.

\bibitem[Mott(1969)]{Mott_1969}
N.~F. Mott.
\newblock Conduction in non-crystalline materials.
\newblock \emph{The Philosophical Magazine: A Journal of Theoretical Experimental and Applied Physics}, 19\penalty0 (160):\penalty0 835--852, 1969.
\newblock \doi{10.1080/14786436908216338}.
\newblock URL \url{https://doi.org/10.1080/14786436908216338}.

\bibitem[Orlovskaya and Browning(2004)]{Orlovskaya2004MixedIE}
Nina~A. Orlovskaya and Nigel~D. Browning.
\newblock \emph{Mixed ionic electronic conducting perovskites for advanced energy systems}.
\newblock 2004.
\newblock URL \url{https://api.semanticscholar.org/CorpusID:92781181}.

\bibitem[Pavarini et~al.(2017)Pavarini, Koch, Scalettar, and Martin]{Pavarini_2017}
Eva Pavarini, Erik Koch, Richard Scalettar, and Richard Martin, editors.
\newblock \emph{{T}he {P}hysics of {C}orrelated {I}nsulators, {M}etals, and {S}uperconductors}, volume~7 of \emph{Schriften des Forschungszentrums Jülich. Reihe Modeling and Simulation}.
\newblock Forschungszentrum Jülich GmbH Zentralbibliothek, Verlag, Jülich, Sep 2017.
\newblock ISBN 978-3-95806-224-5.
\newblock URL \url{https://juser.fz-juelich.de/record/837488}.

\bibitem[Sato et~al.(2009)Sato, Matsuo, Kindo, Kobayashi, and Asai]{Sato_2009}
Keisuke Sato, Akira Matsuo, Koichi Kindo, Yoshihiko Kobayashi, and Kichizo Asai.
\newblock Field induced spin-state transition in lacoo3.
\newblock \emph{Journal of the Physical Society of Japan}, 78\penalty0 (9):\penalty0 093702, 2009.
\newblock \doi{10.1143/JPSJ.78.093702}.
\newblock URL \url{https://doi.org/10.1143/JPSJ.78.093702}.

\bibitem[V. et~al.(2015)V., Kumar, Singh, Gupta, Rayaprol, and Dogra]{AswinV_2015}
Aswin V., Pramod Kumar, Pooja Singh, Anurag Gupta, S.~Rayaprol, and Anjana Dogra.
\newblock Influence of al doping in lacoo3 on structural, electrical and magnetic properties.
\newblock \emph{Journal of Materials Science}, 50\penalty0 (1):\penalty0 366--373, Jan 2015.
\newblock ISSN 1573-4803.
\newblock \doi{10.1007/s10853-014-8595-3}.
\newblock URL \url{https://doi.org/10.1007/s10853-014-8595-3}.

\bibitem[Wenk and Bulakh(2016)]{Wenk_Bulakh_2016}
Hans-Rudolf Wenk and Andrey Bulakh.
\newblock \emph{Minerals: Their Constitution and Origin}.
\newblock Cambridge University Press, 2 edition, 2016.

\bibitem[Yu et~al.(2004)Yu, Wang, Wehrenberg, and Guyot-Sionnest]{Yu_2004}
Dong Yu, Congjun Wang, Brian~L. Wehrenberg, and Philippe Guyot-Sionnest.
\newblock Variable range hopping conduction in semiconductor nanocrystal solids.
\newblock \emph{Phys. Rev. Lett.}, 92:\penalty0 216802, May 2004.
\newblock \doi{10.1103/PhysRevLett.92.216802}.
\newblock URL \url{https://link.aps.org/doi/10.1103/PhysRevLett.92.216802}.

\end{thebibliography}

\end{document}